\documentclass[prb,showpacs,twocolumn,aps]{revtex4}
\usepackage[dvipdfm]{graphicx}
\usepackage[dvipdfm]{color}
\usepackage{dcolumn}
\usepackage{amsmath}
\usepackage{amssymb}

\newcommand{\lnsrnio}{$\rm Ln_{2- \it x}Sr_{\it x}NiO_{4+\delta}$}

\newcommand{\lsno}{$\rm La_{1.67}Sr_{0.33}NiO_{4}$}
\newcommand{\psno}{$\rm Pr_{1.67}Sr_{0.33}NiO_{4}$}
\newcommand{\nsno}{$\rm Nd_{1.67}Sr_{0.33}NiO_{4}$}

\newcommand{\lstno}{$\rm La_{1.725}Sr_{0.275}NiO_{4}$}

\newcommand{\lsxno}{$\rm La_{2- \it x}Sr_{\it x}NiO_{4}$}

\newcommand{\nsxno}{$\rm Nd_{2- \it x}Sr_{\it x}NiO_{4}$}
\newcommand{\rsxno}{$\rm Ln_{2- \it x}Sr_{\it x}NiO_{4}$}

\newcommand{\lsxnod}{$\rm La_{2- \it x}Sr_{\it x}NiO_{4+\delta}$}

\newcommand{\rsxnod}{$\rm Ln_{2- \it x}Sr_{\it x}NiO_{4+\delta}$}

\newcommand{\lno}{$\rm La_{2}NiO_{4}$}

\newcommand{\lsco}{$\rm     La_{2- \it x}               Sr_{\it x}  Cu                          O_{4}$}
\newcommand{\lbco}{$\rm     La_{2- \it x}  Ba_{\it x}               Cu                          O_{4}$}

\newcommand{\lscob}{$\bf    La_{2- \it x}               Sr_{\it x}  Cu                          O_{4}$}

\newcommand{\lcod}{$\rm     La_{2}                                  Cu                          O_{4+\delta}$}
\newcommand{\lstco}{$\rm    La_{1.88}                   Sr_{0.12}   Cu                          O_{4}$}
\newcommand{\lbsco}{$\rm    La_{1.875} Ba_{0.125- \it x}Sr_{\it x}  Cu                          O_{4}$}

\newcommand{\sus}{susceptibility}

\newcommand{\tco}{$T_{\rm CO}$}
\newcommand{\tso}{$T_{\rm SO}$}
\newcommand{\tht}{$T_{\rm HT}$}
\newcommand{\tlt}{$T_{\rm LT}$}

\begin{document}

\title{Unidirectional Diagonal Order and 3D Stacking of Charge Stripes in Orthorhombic $\bf Pr_{1.67}Sr_{0.33}NiO_4$ and $\bf Nd_{1.67}Sr_{0.33}NiO_4$.}
\author{M. H\"ucker$^{1}$, M. v. Zimmermann$^{2}$, R. Klingeler$^{3}$,
        S. Kiele$^{3}$, J. Geck$^{3}$, S. N. Bakehe$^{4}$, J. Z. Zhang$^{1,5}$, J. P. Hill$^{1}$, A. Revcolevschi$^{6}$,
        D. J. Buttrey$^{7}$, B. B\"uchner$^{3}$, and J. M. Tranquada$^{1}$}
\affiliation{$^{1}$Condensed Matter Physics and Material Science
Department, Brookhaven National Laboratory, Upton, New York 11973}
\affiliation{$^{2}$Hamburger Synchrotronstrahlungslabor HASYLAB at
Deutsches Elektronen-Synchrotron, 22603 Hamburg, Germany}
\affiliation{$^{3}$Leibniz-Institute for Solid State and Materials
Research IFW Dresden, 01171 Dresden, Germany} \affiliation{$^{4}${\it
II}.~Physikalisches Institut, Universit\"at zu K\"oln, 50937 K\"oln,
Germany} \affiliation{$^{5}$Cornell University, Ithaca, New York 14850}
\affiliation{$^{6}$Laboratoire de Physico-Chimie de l'Etat Solide,
Universit\'e Paris-Sud, 91405 Orsay Cedex, France}
\affiliation{$^{7}$Department of Chemical Engineering, University of
Delaware, Newark, Delaware 19716}

\date{\today}

\begin{abstract}
The interplay between crystal symmetry and charge stripe order in \psno\
and \nsno\ has been studied by means of single crystal x-ray diffraction.
In contrast to tetragonal \lsno , these crystals are orthorhombic. The
corresponding distortion of the $\rm NiO_2$ planes is found to dictate the
direction of the charge stripes, similar to the case of diagonal spin
stripes in the insulating phase of \lsco . In particular, diagonal stripes
seem to always run along the short $a$-axis, which is the direction of the
octahedral tilt axis. In contrast, no influence of the crystal symmetry on
the charge stripe ordering temperature itself was observed, with $T_{\rm
CO}\sim$240~K for La, Pr, and Nd. The coupling between lattice and stripe
degrees of freedom allows one to produce macroscopic samples with
unidirectional stripe order. In samples with stoichiometric oxygen content
and a hole concentration of exactly 1/3, charge stripes exhibit a
staggered stacking order with a period of three $\rm NiO_2$ layers,
previously only observed with electron microscopy in domains of mesoscopic
dimensions. Remarkably, this stacking order starts to melt about 40~K
below $T_{\rm CO}$. The melting process can be described by mixing the
ground state, which has a 3-layer stacking period, with an increasing
volume fraction with a 2-layer stacking period.\end{abstract}

\pacs{74.72.Dn, 74.25.Ha, 61.12.-q}

\maketitle

\section{Introduction}
In recent years the nickelate \lsxnod\ has been studied intensively
because of its similarity to the high temperature superconductor \lsco
.~\cite{Tranquada98e,Chen93aN,Kajimoto03aN,Ishizaka03aN,Boothroyd03aN,Homes03aN}
Of particular interest is that, in both materials, incommensurable spin
correlations and lattice modulations are observed which are consistent
with the concept of charge and spin stripes.~\cite{Tranquada95a} Undoped,
both systems are insulating antiferromagnets (AF). Doping with Sr ($x$) or
O ($\delta$) introduces hole-like charge carriers, at a concentration
$p=x+2\delta$, into the $\rm NiO_2$ and $\rm CuO_2$ planes, which leads to
a suppression of the commensurate AF order. However, with increasing hole
concentration both systems show a strong tendency towards a frustrated
electronic phase separation.~\cite{Emery93} For certain compositions the
holes segregate into one-dimensional charge stripes, separating
intermediate spin stripes with low hole concentration.~\cite{Tranquada95a}
In \lsxno , stripe order results in an insulating ground
state.~\cite{Cheong94a} Stripes run diagonally to the $\rm NiO_2$ square
lattice and are most stable for $x=0.33$, with the charges ordering at
$T_{\rm CO}\sim$240~K and the spins at $T_{\rm SO}
\sim$190~K.~\cite{Cheong94a,Tranquada95a,Yoshizawa00aN,Kajimoto03aN}
Stripes can be identified by various
techniques.~\cite{Chen93aN,Tranquada95a,Vigliante97a,Yoshinari99aN,Lee02aN,
Li03aN,Blumberg98aN,Abbamonte05a,Langeheine05aN}
Here we focus on the characterization of the charge stripes with x-ray
diffraction by probing the lattice modulation associated with the spatial
modulation of the charge density.~\cite{Vigliante97a}

We are particularly interested in the response of the charge stripe order
to changes of the crystal lattice symmetry. For certain cuprates it was
shown that lattice distortions can pin stripes or influence their
orientation with respect to the lattice. The most prominent example is
observed in \lbco\ and Nd or Eu-doped \lsco\ around a hole doping of
$x=1/8$, where stripes are pinned parallel to the $\rm CuO_2$ square
lattice as a result of a structural transition from orthorhombic to
tetragonal symmetry.~\cite{Fujita04a,Tranquada95a,Niemoeller99a,Klauss00a}
In samples with a fully developed static stripe order, superconductivity
is strongly suppressed.~\cite{Wagener97a,Tranquada97a,Klauss00a,Fujita04a}
Another example is observed in orthorhombic \lsco\ below the metal
insulator (MI) transition ($x<0.06$), i.e., in the short-range ordered
spin-stripe phase. The ground state is insulating, with spin stripes
running diagonally to the square lattice.~\cite{Wakimoto99a,Fujita02a} In
this case the coupling to the orthorhombic distortions is evident from the
finding that the spin stripes always order parallel to the orthorhombic
$a$-axis.~\cite{Wakimoto00a} Note that, below the MI transition, no
diffraction evidence for charge stripes has been found. Therefore, one has
to consider the possibility that here the incommensurate magnetism might
emerge from ground states different from stripe order, such as helical
spin structures or disorder-induced spin
topologies.~\cite{Shraiman92a,Gooding97,Berciu04a,Sushkov04a,Luescher05a,Lindgard05a}

To gain insight into the mechanisms of the charge-lattice coupling in the
cuprates, we have carried out a study of the diagonal stripes in the
isostructural nickelates. The particular question which motivated this
work is: do lattice distortions in the nickelates, which are very similar
to those in the cuprates, have an impact on the stripe orientation? In
this context, we consider the crystal structure of \rsxnod\ with Ln=La,
Pr, and Nd.

\begin{figure}[t]
\center{\includegraphics[width=1\columnwidth,angle=0,clip]{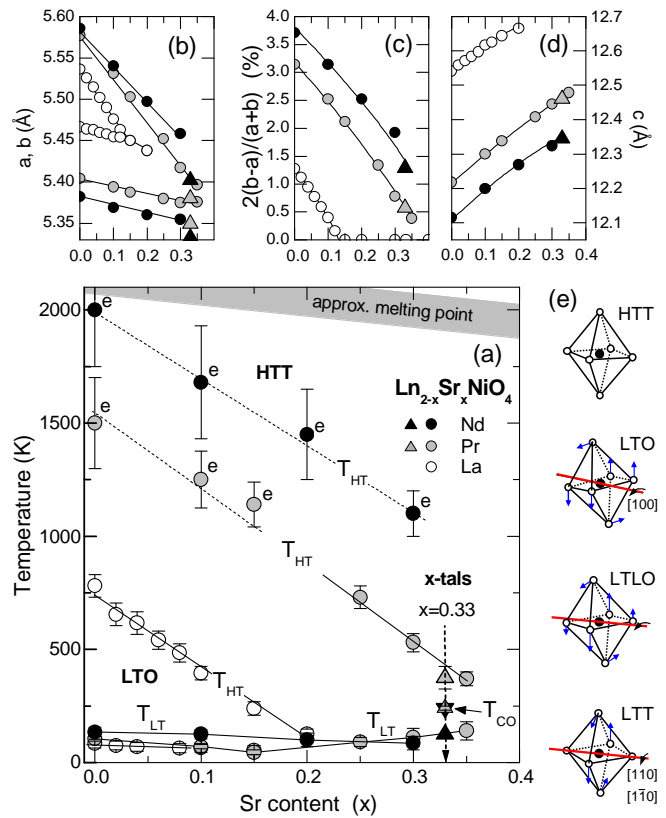}}
\caption[]{(Color online) (a) Phase diagram of \rsxno\ with Ln=La, Pr, and
Nd as obtained from x-ray powder diffraction on polycrystalline
specimens.~\cite{BakeheDiss,FriedtDipl} For Pr and Nd the HTT/LTO phase
boundary is shifted to higher $x$ and $T$ with respect to La. Data points
marked with an "e" are linear extrapolations of measurements up to 800~K.
Top: Lattice parameters $a$, $b$, and $c$ as well as orthorhombic strain
at room temperature. Data for the single crystals at $x=0.33$ are
indicated by triangular symbols. (e) Tilt directions of the $\rm NiO_6$
octahedra in the different structural phases.} \label{fig1}
\end{figure}

As summarized in Fig.~\ref{fig1}(a), undoped \lno\ (open symbols)
undergoes two structural transitions: From the high-temperature-tetragonal
(HTT) phase to the low-temperature-orthorhombic (LTO) phase at $\rm
T_{HT}\simeq 750$~K, and from the LTO phase to the low-temperature
less-orthorhombic (LTLO) phase at $\rm T_{LT}\simeq 80$~K. (See
Sec.~\ref{structure} for definition of phases.) With increasing Sr
content, both transitions decrease in temperature and, above a critical Sr
content of $x_c\simeq 0.2$, the HTT phase is stable for all
temperatures.~\cite{Rodriguez88aN,Medarde97aN,Sachan95aN,FriedtDipl,Huecker04a}
Hence, in \lsno , stripe order takes place in a tetragonal lattice, which
results in an equal population of domains with stripes running along the
orthorhombic [100] and [010] directions.~\cite{Boothroyd03aN,Woo05aN}
Obviously, such stripe twinning complicates any anisotropy study of
stripes, such as electronic transport parallel and perpendicular to the
charge stripes or magnetic excitations parallel and perpendicular to the
spin stripes.

The substitution of trivalent La with the smaller, isovalent Pr or Nd
causes a significant increase of the chemical pressure on the $\rm NiO_2$
planes.~\cite{Huecker04a,Ishizaka03aN} As a result, the HTT/LTO phase
boundary shifts to much higher temperatures and Sr concentrations so that
,even for $x=0.33$, the transition takes place above room temperature (see
Fig.~\ref{fig1}). Therefore, the Pr and the Nd-based systems allow us to
investigate the formation of stripes under the influence of orthorhombic
strain. The effect of the Pr and Nd substitution on both the high
temperature and the low temperature structural transitions has been
characterized by several groups, although in some studies samples with
non-stoichiometric oxygen content were
investigated.~\cite{Martinez91aN,Buttrey90aN,Medarde94aN} In
Ref.~\onlinecite{Ishizaka03aN}, the charge order in \nsxno\ with
$x\geqslant 0.33$ was studied with the focus on the evolution from stripe
to checkerboard order.

In the present paper we focus on the coupling between charge stripes and
lattice distortions in the LTO phase of \psno\ and \nsno . In particular,
we find that stripes always order parallel to the short $a$-axis of the
LTO unit cell. This will allow one to produce macroscopic samples with
unidirectional stripe order by detwinning the crystals under
unidirectional strain. The identified stripe orientation is the same as
for the short range spin stripe order in lightly doped \lsco\ with LTO
structure, as will be discussed. The critical temperature, \tco , turns
out to be the same as for \lsno , i.e., no significant dependence on the
orthorhombic strain is observed. Furthermore, we find a strong dependence
of the stacking order of stripes along the $c$-axis on the oxygen and hole
concentration. Notably, crystals containing excess oxygen exhibit strong
stacking disorder. Crystals with stoichiometric oxygen content show a
tendency towards a 3-layer stacking period, resulting in a unit cell
enlarged by a factor of 1.5 along the $c$-direction. However, only for a
hole concentration of exactly 1/3 does this 3-layer period lead to a well
defined superstructure.

\section{Crystal structure}
\label{structure}
\lnsrnio , with $\rm Ln=La$, Pr, or Nd, crystallizes in the $\rm K_2NiF_4$
structure and at high temperature is expected to assume tetragonal
symmetry, space group $I4/mmm$, regardless of the Sr and O
contents.~\cite{Tamura96aN,Sullivan91aN} For convenience, it is common to
index all phases on the basis of the orthorhombic $\sqrt{2}a \times
\sqrt{2}b \times c$ supercell, in which case the space group of the HTT
phase is $F4/mmm$.~\cite{Huecker04a}
As a function of temperature and doping, a total of four structural phases
are observed, which can be described by different buckling patterns of the
$\rm NiO_6$ octahedral network.~\cite{Axe89,Crawford91,Huecker04a} The HTT
phase is the only phase with untilted octahedra, i.e., the $\rm NiO_2$
planes are flat [cf. Fig.~\ref{fig1}(e)]. In the LTO phase, space group
$Bmab$, octahedra rotate about the $a$-axis, which runs diagonally to the
$\rm NiO_2$ square lattice. In the LTT phase, space group $P4_2/ncm$, the
octahedral tilt axis is parallel to the Ni-O bonds, which means that it
has rotated by $\alpha = 45^\circ$ with respect to the LTO phase.
Moreover, its direction alternates between [110] and [1-10] from plane to
plane. The LTLO phase, space group $Pccn$, is an intermediate between LTO
and LTT with $\alpha < 45^\circ$.

\section{Experiment}
Two single crystals, with compositions \psno\ and \nsno\ and sizes of
4.5~mm in diameter and several centimeters in length, were grown at the
Laboratoire de Physico-Chimie des Solides at Orsay by the
travelling-solvent floating-zone method. In both cases, crystal growth was
performed in 1 atm of air.
The polycrystals used to map the phase diagram in Fig.~\ref{fig1} were
synthesized by a standard solid-state
reaction.~\cite{BakeheDiss,FriedtDipl} To remove non-stoichiometric oxygen
interstitials (excess oxygen), the polycrystals were annealed under
reducing conditions that depended on the Sr content. Powder x-ray
diffraction patterns of the polycrystals for temperatures up to 800~K were
taken with a standard laboratory
diffractometer.~\cite{BakeheDiss,FriedtDipl}

Single crystal x-ray diffraction experiments were performed at beamline
X22C of the National Synchrotron Light Source (NSLS) at Brookhaven and at
beamline BW5 of the Hamburg Synchrotron Laboratory (HASYLAB). At BW5, the
photon energy was set to 100~keV.~\cite{Bouchard98} At this energy, a
sample several millimeters thick can be studied in transmission geometry,
allowing one to probe its bulk. In contrast, at Brookhaven 8.1 keV
photons, which have a penetration depth on the order of $10~\mu$m, were
used, so that, here, samples were studied in reflection geometry. The
\psno\ crystal, with a polished and twinned [100]/[010] surface, at first
has been studied as-grown-in-air at X22C and BW5. Subsequently, it was
studied again at X22C, after removing the excess oxygen ($\delta =0$) by
Ar-annealing at 900$^\circ$C for 24~h. The \nsno\ crystal was studied at
BW5 after being subjected to an identical Ar-annealing. For the hard x-ray
diffraction experiments no polished surface is required.

The specific heat of the Ar-annealed \psno\ crystal was measured with a
Physical Property Measurement System from Quantum Design. The $ab$-plane
resistivity of a bar-shaped, Ar-annealed \nsno\ crystal was measured using
the four-probe technique.

\section{Results}
\subsection{Phase diagram}
To put the results for the two single crystals ($x=0.33$) into context
with \lsno , we refer back to Fig.~\ref{fig1}(a), the Sr-doping phase
diagram for Ln=La, Pr, and Nd.~\cite{BakeheDiss,FriedtDipl} The shift of
the HTT/LTO phase boundary to higher temperatures $T_{\rm HT}$ and $x$
with decreasing rare earth ionic radius is obvious ($\rm La^{\rm
3+}>Pr^{\rm 3+}>Nd^{\rm 3+}$). In those cases where $T_{\rm HT}>800$~K,
the transition temperature was determined by linearly extrapolating the
orthorhombic strain to higher temperatures. The low-temperature structural
transition was observed in all orthorhombic samples. However, it changes
qualitatively as a function of the Sr content. At low $x$, the transition
is of the discontinuous LTO/LTLO type whereas, at higher $x$, it becomes
continuous. No reliable information was obtained for the crossover
concentrations, or whether these are different for the La, Pr, and
Nd-based systems.
However, with $x$ approaching $x_c(T_{\rm HT}=0)$, the transition becomes
very broad and the orthorhombicity barely decreases below \tlt . For
Nd-based samples, similar results were reported in
Ref.~\onlinecite{Medarde94aN}. Note that while in the Pr-based polycrystal
with $x=0.35$ a weak LTO/LTLO is still visible, it is not observed in the
Pr-based single crystal with $x=0.33$ (see next section). In
Figs.~\ref{fig1}(b) and (d) we show the Sr-doping dependence of the
lattice parameters, and in Fig.~\ref{fig1}(c) the orthorhombic strain at
room temperature.

Corresponding data for the single crystals (triangular symbols) are in
fair agreement with the polycrystal data (cf. Fig.~\ref{fig1}). The
largest deviations concern the absolute values of the lattice parameters
$a$ and $b$. It seems that the single crystal data are systematically too
low by about 0.5\%. Differences of this magnitude are within the error of
our single crystal diffraction experiment, where $a$, $b$, and $c$ were
determined from basically one reflection each. The temperatures of the
HTT/LTO transition are 370~K (315~K) for the surface (bulk) in \psno\
(details in Sec.~\ref{unidirect}), and an estimated 1100~K in \nsno . In
addition, the Nd-based crystal undergoes the LTO/LTLO transition at
$\sim$125~K.

\begin{figure}[t]
\center{\includegraphics[width=0.95\columnwidth,angle=0,clip]{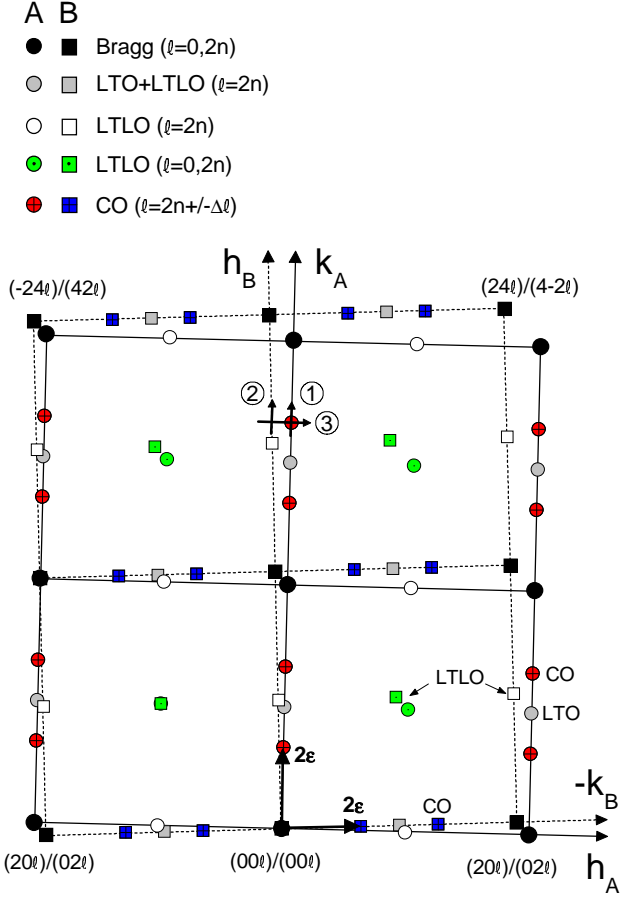}}
\caption[]{(Color online) Projection along $l$ of the $(h,k)$-zone. Domain
$A$: round symbols. Domain $B$: quadratic symbols. Tilt between the two
domains exaggerated. Fundamental reflections are black, LTO superlattice
peaks grey, LTLO superlattice peaks green (with dot) and white. The charge
stripe peaks (CO) are red and blue (with cross), and appear at
incommensurate positions, displaced by $2\epsilon$ along $k$ from Bragg
peaks. Reflection conditions for $l$ are indicated in the legend. $\Delta
l$ depends on the stacking order, which we discuss in
Sec.~\ref{stripestacking}. The three enumerated scans are those in
Fig.~\ref{fig3}(c,d).} \label{fig2}
\end{figure}

\begin{figure}[t]
\center{\includegraphics[width=0.8\columnwidth,angle=0,clip]{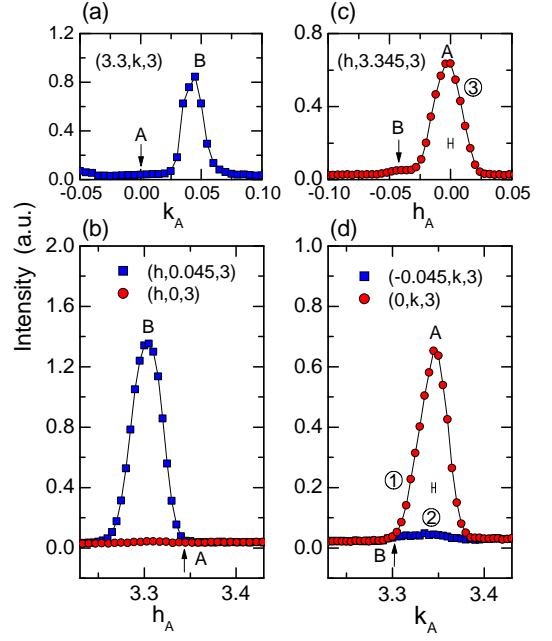}}
\caption[]{(Color online) Scans performed on \nsno\ at $T= 200$~K through
possible charge stripe peak positions ($0,4-2\epsilon ,3$) and
($4-2\epsilon, 0 ,3$) of domains $A$ (round symbols) and $B$ (quadratic
symbols) in the orientation matrix of domain $A$. Arrows indicate expected
peak positions in the case that stripes would also run along the $b$-axis.
Enumerated scans in (c,d) are indicated in Fig.~\ref{fig2}. The H-signs in
(c) and (d) indicate the instrumental resolution full width.} \label{fig3}
\end{figure}

The approximate melting temperatures in Fig.~\ref{fig1}(a) are based on
measurements of the surface temperature of the melt using an optical
pyrometer. These results were obtained in earlier crystal growth
experiments on \lnsrnio\ with $x=0$, 0.25, and 1.5 in which the skull
melting method was applied. We mention that our results in Fig.~\ref{fig1}
are consistent with available data by other groups, in particular when
taking the effect of oxygen doping into
account.~\cite{Martinez91aN,Medarde94aN,Sullivan91aN,Ishizaka03aN,Huecker04a}
However, Fig.~\ref{fig1} may be the most coherent account of \lnsrnio\
with stoichiometric oxygen content for $x \leq 0.35$ and Ln=La, Pr, and
Nd.

\begin{figure}[t]
\center{\includegraphics[width=0.63\columnwidth,angle=0,clip]{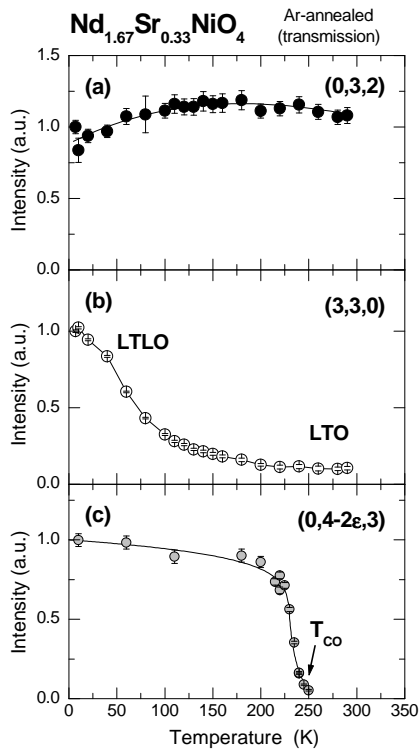}}
\caption[]{Temperature dependence of the normalized intensities of (a) the
LTO superstructure peak, (b) the LTLO superstructure peak and (c) the
charge stripe order peak for the Ar-annealed \nsno\ crystal. Data
collected transmission geometry (100~keV).} \label{fig4}
\end{figure}

\subsection{Structural transitions in single crystals}
In this and the next section, we focus on the temperature dependence of
the crystal symmetry and its influence on charge stripe order. As
mentioned earlier, the Pr and Nd-based compounds undergo the HTT/LTO
transition above room temperature. It is well known that below this
transition, crystals tend to form twin boundaries in the $ab$-plane. In
particular, in space group $Bmab$, up to four twin domains can form,
resulting in a corresponding manifold of fundamental
reflections.~\cite{Wakimoto00a} In both crystals studied, only two out of
these four domains are present, which we call domains $A$ and $B$.

Figure~\ref{fig2} shows a projection of the $(hk)$-zone along the
$l$-direction in the case of the LTLO phase with charge stripe order, as
we find it in the \nsno\ crystal at low temperatures. The legend shows the
$l$-conditions for the different types of reflections. The fundamental
reflections of the domains $A$ and $B$ are indicated by black symbols. The
angle between the two domains amounts to $\sim$0.7$^\circ$ at 200~K
(exaggerated in the figure). Superstructure reflections in the LTO phase,
such as (032), are indicated by grey symbols. In the LTLO phase,
additional reflections appear, such as (110) and (302), which are
indicated by symbols with dot (green) and open symbols, respectively.
Furthermore, the angle between the domains is smaller than in the LTO
phase. In the LTT phase, the reflections of the two domains are merged
into single peaks. In our \nsno\ crystal, the LTT phase is not observed
down to $T=7$~K.

\begin{figure}[t]
\center{\includegraphics[width=0.75\columnwidth,angle=0,clip]{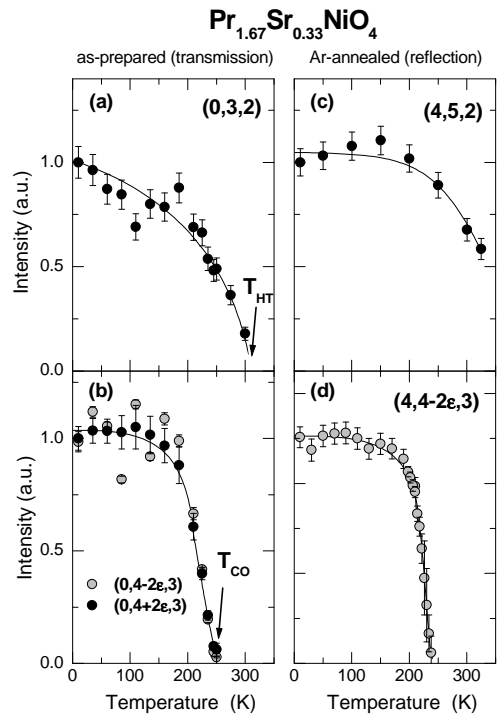}}
\caption[]{Temperature dependence of the normalized intensities of (a,c)
the LTO superstructure peak and (b,d) the charge stripe order peak in
\psno . Data on left side in transmission geometry (100~keV) on
as-prepared-in-air crystal, and on right side in reflection geometry
(8.1~keV) after Ar-annealing.} \label{fig5}
\end{figure}

\subsection{Unidirectional Stripes in the LTO phase}
\label{unidirect}
Now let us turn to the weak reflections which appear below $T_{\rm CO}\sim
240$~K due to the charge stripe order, indicated in Fig.~\ref{fig2} by
symbols with cross for domain $A$ (red) and domain $B$ (blue). In both
crystals, as well as in each of their two domains, these peaks were always
found along $k$. This conclusion is based on scans through possible
charge-peak positions in both domains, similar to those presented in
Fig.~\ref{fig3}.~\cite{NSNOpeaks} Note that all scans where performed in
the orientation matrix of domain $A$, so that the positions of the peaks
of domain $B$ appear slightly shifted. The results indicate a lattice
modulation along the $b$-axis, which means that in the LTO phase stripes
are parallel to the short $a$-axis. Obviously, the orthorhombic strain in
the LTO phase dictates the direction of the stripes. No change of the
stripe pattern was observed in the LTLO phase of \nsno , which is not
surprising because in the LTLO phase one still has $a<b$.

For the \nsno\ crystal, the temperature dependence of various
superstructure peaks is shown in Fig.~\ref{fig4}. Below room temperature,
the (032) peak intensity first increases slightly, but starts to decrease
below $\sim$125~K. The decrease of the (032) peak coincides with a strong
increase of the (330) peak, indicating the LTO/LTLO
transition.~\cite{LTLOpeak} Above $\sim$125~K, the (330) peak is very weak
and broad but remains visible up to room temperature.~\cite{newXTAL} The
decrease of the (032) peak in the LTLO phase is due to a shift of
intensity to the (302) peak (white symbols in Fig.~\ref{fig2}). Below the
charge stripe transition at $T_{\rm CO} \simeq 240$~K, the intensity of
the $(0,4-2\epsilon,3)$ peak first increases steeply, but tends to
saturate below 200~K. There is no obvious change of the intensity of this
peak due to the LTO/LTLO transition.

\begin{figure}[t]
\center{\includegraphics[width=0.7\columnwidth,angle=0,clip]{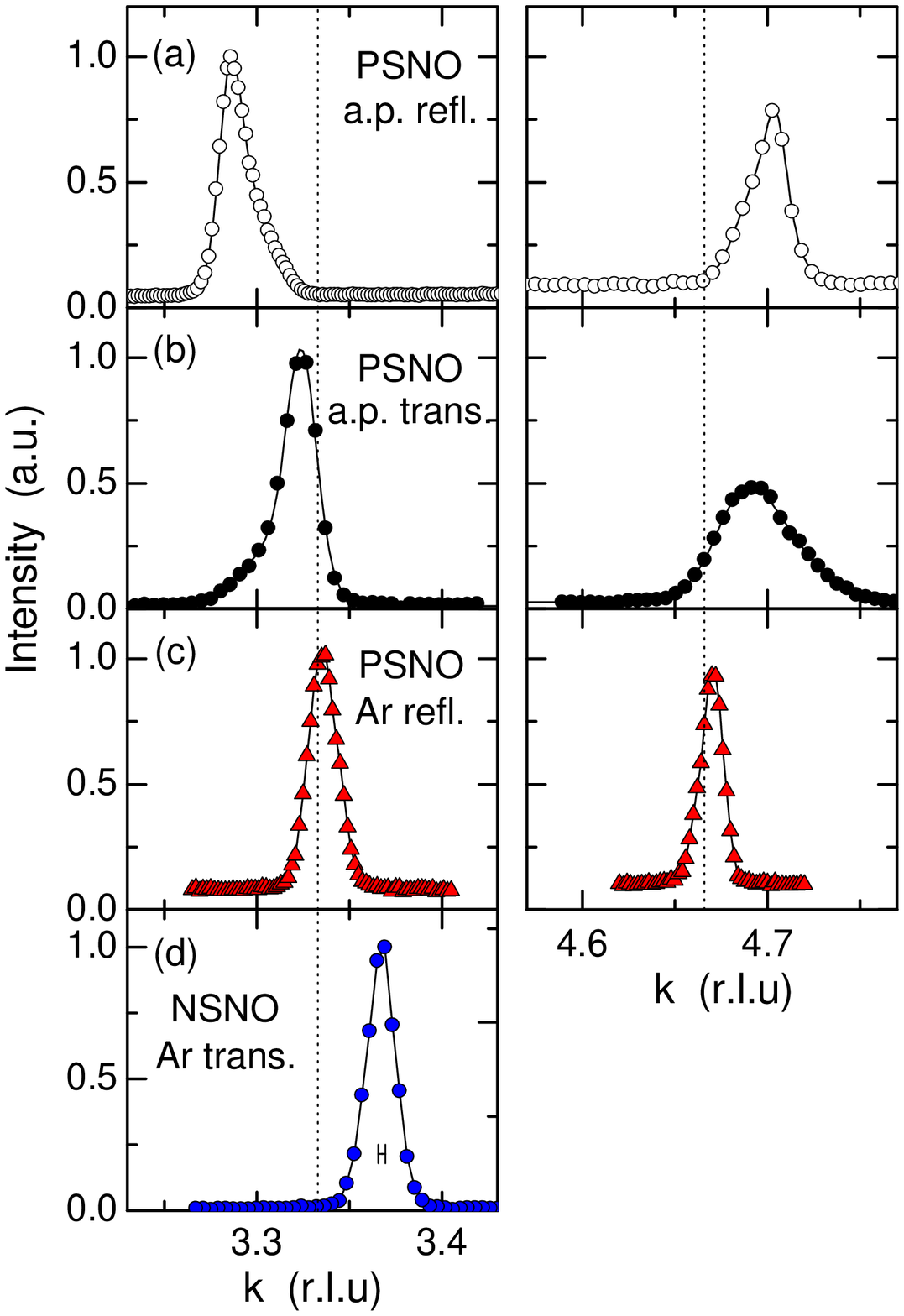}}
\caption[]{(Color online) Comparison of $k$-scans through $\rm
(h,4-2\epsilon,3)$ and $\rm (h,4+2\epsilon,3)$ at 10~K for \psno\ (PSNO)
(a-c) and \nsno\ (NSNO) (d). Scans in (c) performed at $h=4$ (triangular
symbols). All other scans at $h=0$. Data collected (a) in reflection and
(b) in transmission on as-prepared sample, (c) in reflection and (d) in
transmission on Ar-annealed samples. The intensities of the peaks at
$k=4+2\epsilon$ were normalized by same factors as their counterparts at
$k=4-2\epsilon$. The H-shaped sign in (d) indicates typical instrumental
resolution full width.} \label{fig6}
\end{figure}

In Fig.~\ref{fig5} we show corresponding data for \psno . As mentioned
before, no LTO/LTLO transition was observed in this crystal. The figures
on the left side show 100~keV x-ray data collected in transmission
geometry on the crystal in the as-prepared-in-air state ($\delta \gtrsim
0$). The (032) reflection strongly decreases with increasing temperature,
indicating a LTO/HTT transition temperature of $\sim$315~K. Charge stripe
order sets in at the same temperature as in \nsno , although the
transition is somewhat broader [Fig.~\ref{fig5}(b)]. The latter effect is
due to the non-stoichiometric oxygen and its inhomogeneous distribution in
the crystal. The influence of excess oxygen will be discussed in more
detail in Secs.~\ref{stripedistance} and \ref{stripestacking}.

The figures on the right side of Fig.~\ref{fig5} show data collected at
8.1~keV in a reflection geometry after Ar-annealing ($\delta \simeq 0$).
The charge stripe transition now is much sharper, though the transition
temperature remains the same. On the other hand, the LTO/HTT transition is
observed at a significantly higher temperature. [Note that in this
experiment we have studied the equivalent (452) reflection.] In addition
to the data in Fig.~\ref{fig5}, we have collected another set of data (not
shown) in reflection geometry on the as-prepared crystal. Also in this
case the LTO/HTT transition occurs at a much higher temperature (370~K)
than in transmission geometry. Therefore, we think that \tht\ not only
depends on the hole concentration [cf. Fig.~\ref{fig1}(a)], but generally
seems to be somewhat higher in the surface layer than in the bulk.

\begin{figure}[t]
\center{\includegraphics[width=0.75\columnwidth,angle=0,clip]{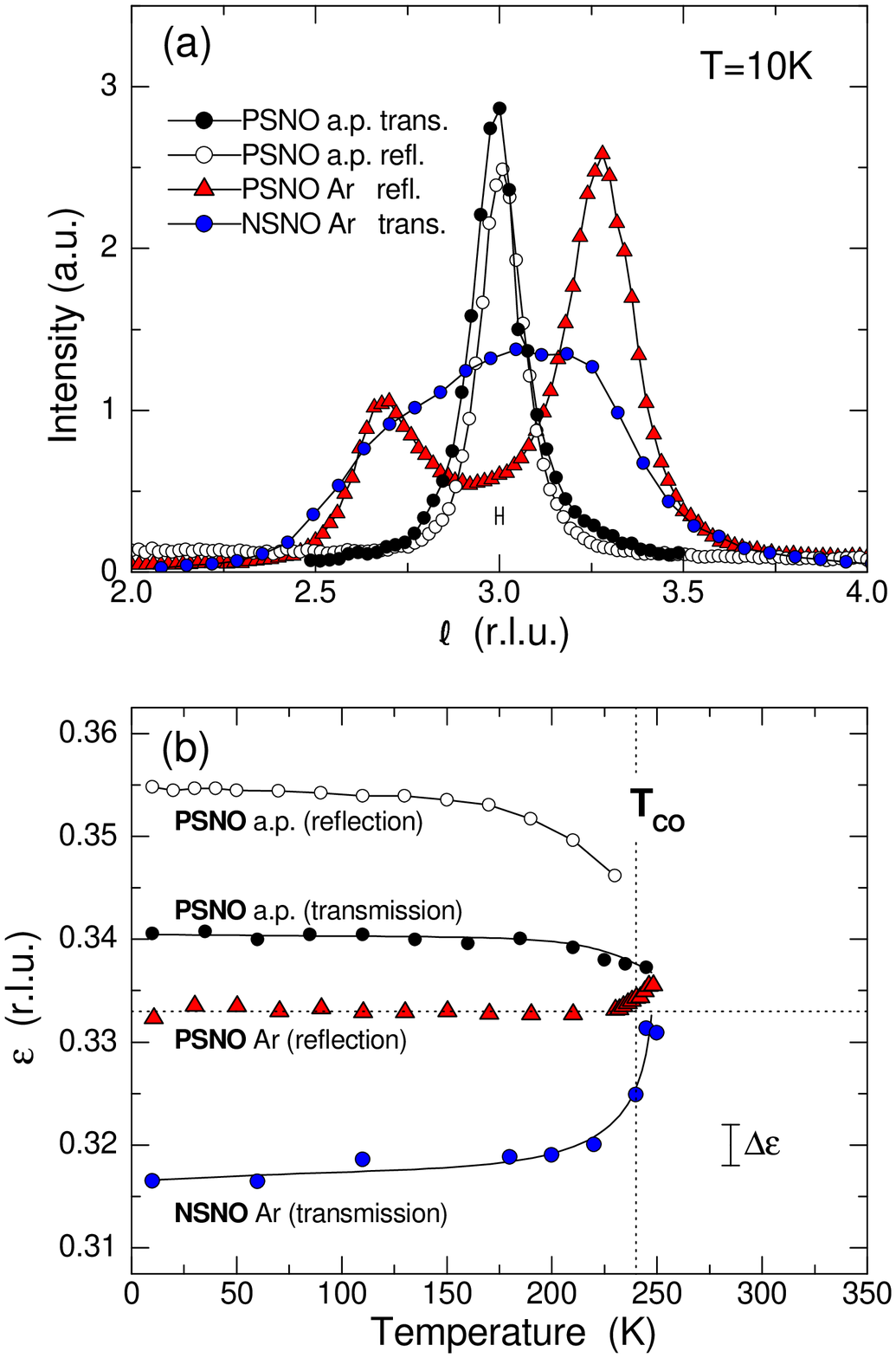}}
\caption[]{(Color online) (a) Comparison of $l$-scans at 10~K through
charge peak at $\rm (h,4-2\epsilon,3)$ for \psno\ and \nsno . For
triangular symbols scans were performed at $h=4$. All other scans at
$h=0$. Typical instrumental resolution full width is indicated by H-sign.
(b) Incommensurability $\epsilon$ as a function of temperature. Error for
$\epsilon$ indicated by vertical bar. Symbols and color code for samples
and experiments the same as in Fig.~\ref{fig6}.} \label{fig7}
\end{figure}

\subsection{In-plane Charge Stripe Distance}
\label{stripedistance}
Excess oxygen $\delta$ increases the total hole concentration
$p=x+2\delta$ in the $\rm NiO_2$ planes. One way to measure $p$ is to
measure the in-plane distance between the charge stripes via the
incommensurability $\epsilon$ (see Fig.~\ref{fig2}). The property
$\epsilon(p)$ was studied intensively in recent years and it is well known
that it is not precisely linear.~\cite{Yoshizawa00aN} However, around
$p=1/3$, one finds that $\epsilon=p$ is approximately satisfied. Hence,
deviations from $p=1/3$ can be probed quite accurately. A precise way to
determine $\epsilon$ is to take one fourth the distance between the
simultaneously measured reflections at $k=2n-2\epsilon$ and
$k=2n+2\epsilon$.

\begin{figure}[t]
\center{\includegraphics[width=0.95\columnwidth,angle=0,clip]{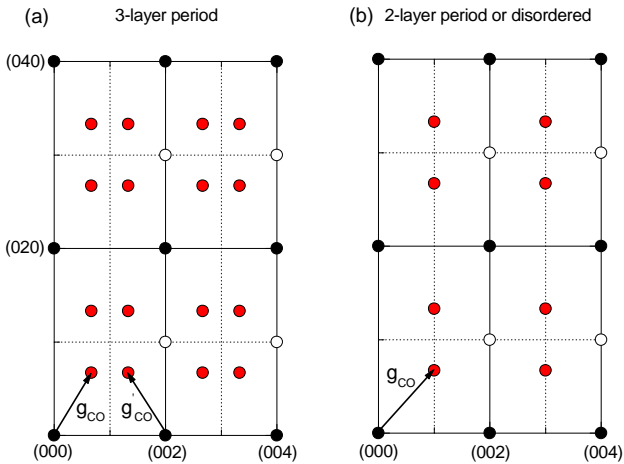}}
\caption[]{(Color online) Diagrams indicating the positions of the charge
stripe peaks (red symbols) as well as corresponding ordering wave vectors
for samples with (a) 3-layer stripe stacking period and (b) 2-layer period
or strong stacking disorder. Full symbols indicate fundamental
reflections, open symbols LTO superstructure reflections. In (b) we assume
a 2-layer period similar to Fig.~\ref{fig10}(c). Additional charge peaks
would appear for $l$ even, assuming the stacking depicted in
Fig.~\ref{fig10}(b).~\cite{evenpeaks}} \label{fig8}
\end{figure}

In Fig.~\ref{fig6} we show representative scans for as-prepared-in-air and
Ar-annealed crystals taken at different photon energies, as indicated in
the figure. The larger the distance between the peaks, the larger $p$. The
dashed lines are for $p=1/3$. By comparing the 8.1~keV and 100~keV data in
Figs.~\ref{fig6}(a) and (b), we find that in the as-prepared \psno\
crystal the oxygen content in the surface is larger than in the bulk. Only
after the Ar-annealing is the hole concentration in this crystal very
close to the nominal value $p=x=1/3$, suggesting that $\delta\simeq 0$
[Fig.~\ref{fig6}(c)]. In the Ar-annealed \nsno\ crystal we observe
$p<1/3$, which we attribute to a Sr content slightly below $x=1/3$,
assuming that $\delta\simeq 0$ [Fig.~\ref{fig6}(d)]. Note that for \nsno\
we have measured the peak at $k=4-2\epsilon$, only. Another feature is
that, in the as-prepared \psno\ crystal, the peaks are asymmetric and
obviously broader in $k$ than in both Ar-annealed crystals, indicating
that excess oxygen leads to a enhanced distribution of $\epsilon$.

In Fig.~\ref{fig7}(b) we show the corresponding temperature dependencies
of $\epsilon$. The symbols are the same as in Fig.~\ref{fig6}. At low
temperatures one can clearly see the deviations from $\epsilon=1/3$ for
as-prepared \psno\ and Ar-annealed \nsno . It is believed that the low
temperature values of $\epsilon$ represent the true hole concentration
$p$. However, close to \tco , in all measurements $\epsilon$ gravitates
towards a value of $1/3$. This lock-in effect is well known and indicates
that the in-plane stripe distance prefers to be three Ni sites (1.5$\times
b$), i.e., commensurate with the
lattice.~\cite{Tranquada96aN,Vigliante97a,Wochner98aN,Kajimoto01aN,Ishizaka04aN}
Note that in the case of the Ar-annealed \psno\ crystal, $\epsilon\simeq
1/3$ for all temperatures, which indicates that this is in fact the only
experiment where we were truly looking at a sample with $p = 1/3$.

\begin{figure}[t]
\center{\includegraphics[width=0.65\columnwidth,angle=0,clip]{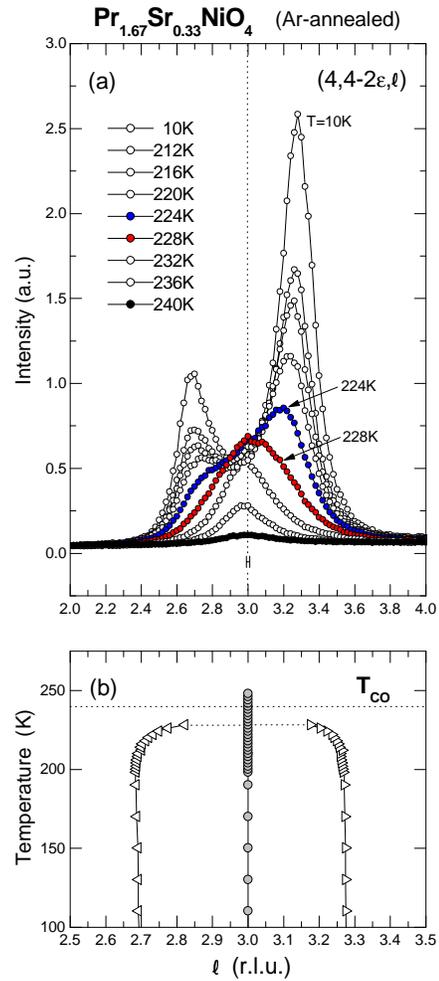}}
\caption[]{(Color online) (a) $l$-scans through $\rm (4,4-2\epsilon,3)$ of
Ar-annealed \psno\ for different temperatures. Instrumental resolution
full width is indicated by small H-sign. (b) Peak positions as a function
of $T$ from fit with three peaks. The position of the central peak was
kept fixed at $l=3$.} \label{fig9}
\end{figure}

\subsection{Three Dimensional Charge Stripe Order}
\label{stripestacking}
The most dramatic effect the Ar-annealing had on the \psno\ crystal was a
drastic change of the correlations between charge stripes along the
$c$-axis. Corresponding $l$-scans through the charge stripe peak at $\sim
10 $~K are shown in Fig.~\ref{fig7}(a). In the as-prepared crystal, a
single narrow peak appears at $l=3$, which is in accordance with results
by other groups on \lsxnod .~\cite{Vigliante97a,Lee01aN,Du00aN} After the
Ar-annealing, however, the peak is split with two pronounced maxima
appearing at $l\simeq 2.7$ and $l\simeq 3.3$. Since these values are close
to $l=2+ 2/3$ and $l=4-2/3$, the split most likely indicates that a long
range stacking order of stripes has developed along the $c$-axis with a
period of $3/2\times c$, corresponding to a period of three $\rm NiO_2$
layers. Figure~\ref{fig8} shows the locations of the charge stripe peaks
in the $(0kl)$-zone as observed for the Ar-annealed (left) and the
as-prepared \psno\ crystal (right). The corresponding ordering wave
vectors are ${\bf g}_{\rm CO}=(0,2\epsilon,2/3)$ and ${\bf g^{\bf '}}_{\rm
CO}=(0,2\epsilon,-2/3)$ as well as ${\bf g}_{\rm CO}=(0,2\epsilon,1)$,
respectively. Details will be discussed in Sec.~\ref{discussion}. In the
\nsno\ crystal, a similarly clear splitting in $l$ was not observed, but,
as one can see in Fig.~\ref{fig7}(a), the scan is extremely broad and
shows two shoulders at comparable $l$-positions, plus a central maximum,
indicating the presence of incommensurate peaks obscured by disorder.

In Fig.~\ref{fig9}(a) we show the temperature dependence of $l$-scans
through $\rm (4,4-2\epsilon,3)$ for \psno . To extract the peak positions
we have fit the data with three lines, keeping the position of the central
line fixed at $l=3$. As one can see in Fig.~\ref{fig9}(b), the splitting
in $l$ is about constant up to 200~K and then starts to decrease.
Eventually, for $T\gtrsim 228$~K the three peaks have merged into a single
reflection at $l=3$, indicating the loss, or a qualitative change, of the
long-range correlations along the $c$-axis.

In Fig.~\ref{fig9} peaks appear at slightly asymmetric positions with
respect to $l=3$. We have performed additional scans through charge and
nearby Bragg peaks, which confirm that the peak positions are symmetric.
The deviations in Fig.~\ref{fig9} are due to a small misalignment of the
crystal, as well as the temperature dependence of the lattice parameter
$c$. In our data analysis in Sec.~\ref{discussion}, this deviation is
compensated by a small offset in $l$.

\begin{figure}[t]
\center{\includegraphics[width=1\columnwidth,angle=0,clip]{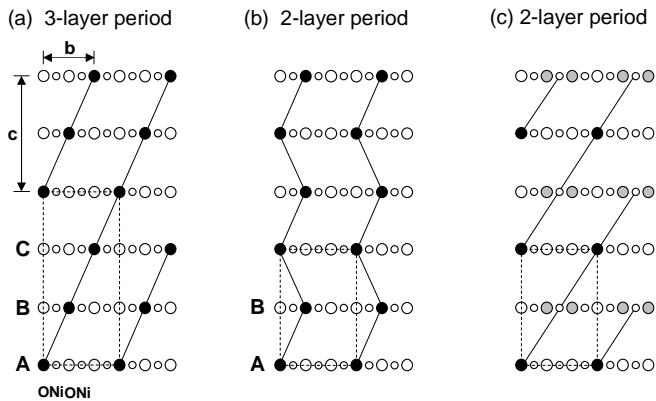}}
\caption[]{Stripe stacking patterns. (a) 3-layer period comparable to
cubic stacking, (b) 2-layer period comparable to hexagonal stacking, (c)
2-layer period due to body centered stacking with alternating layers of
Ni-O-Ni-bond centered (black symbols) and Ni-site centered (grey symbols)
stripes. } \label{fig10}
\end{figure}

\section{Discussion}
\label{discussion}
\subsection{Modelling the stacking of stripes}
Previous studies of \lsxnod\ have shown that the stacking of stripes along
the $c$-axis is mainly controlled by two
mechanisms~\cite{Tranquada96aN,Wochner98aN}: the minimization of the
Coulomb repulsion between the charge stripes, and the interaction of the
charge stripes with the underlying lattice. The Coulomb interaction favors
a body centered-type stacking to maximize the distance between nearest
neighbor stripes in adjacent planes. On the other hand, the interaction
with the lattice favors shifts by increments of half of the in-plane
lattice parameter. This means that nearest-neighbor stripes do not stack
on top of each other. For a hole content of $p=0.25$ and 0.5, a
body-centered stacking also satisfies commensurability with the lattice.
This is not the case for $p=1/3$. Here, a body-centered stacking is
achieved only in the case of an alternation of layers with site-centered
and bond-centered stripes, similar to the stacking shown
Fig.~\ref{fig10}(c). However, our results for Ar-annealed \psno\ strongly
suggest a stripe ground state with a 3-layer period, consistent with a
stacking similar to Fig.~\ref{fig10}(a). Stripes successively shift by
increments of $b/2$, i.e., there are three different types of planes with
stripes at in-plane positions 0, $b/2$ and $b$. A stacking pattern such as
that in Fig.~\ref{fig10}(a) but with all stripes centered on Ni-sites is
possible, too.

\begin{figure}[t]
\center{\includegraphics[width=0.7\columnwidth,angle=0,clip]{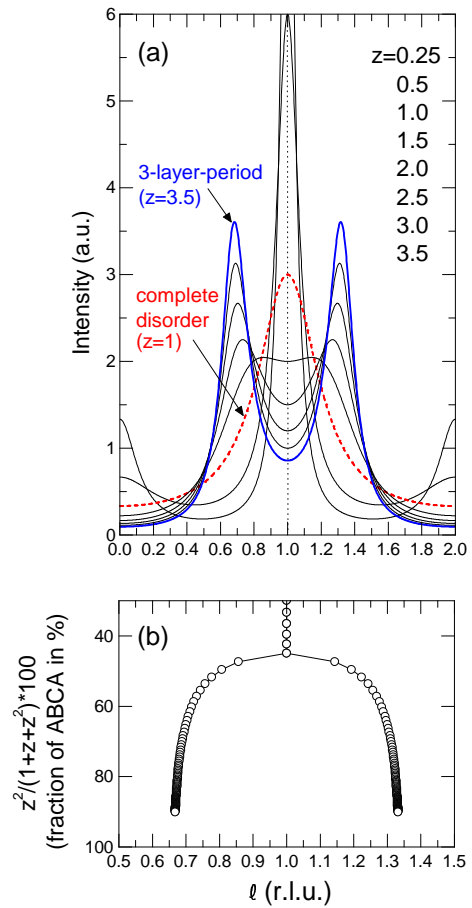}}
\caption[]{(Color online) (a) Calculated scattering intensity for
$l$-scans as a function of the probability $z$, where $z=0$ for hexagonal
close packing and $z=\infty$ for cubic close packing (after
Ref.~\onlinecite{Hendricks42a}). The curve for $z=1$ (dashed) represents
complete disorder. The curve for $z=3.5$ represents a system with a fairly
developed 3-layer stacking period. (b) $l$-positions of the maxima as a
function of the volume fraction of ABCA stacking segments.} \label{fig11}
\end{figure}

This problem can be projected onto that of the stacking of close-packed
layers of atoms, where layers can be shifted by 1/3 of the primitive
translation, again resulting in three types of planes: A, B and
C.~\cite{Hendricks42a} The two most simple arrangements of these planes
are the hexagonal packing ...ABAB...,~\cite{BCBC} and the cubic packing
...ABCABC..., corresponding to stripe stacking patterns with 2-layer and
3-layer periods, respectively [cf. Fig.~\ref{fig10}(a,b)]. Energetic
differences between these two stacking types appear only if
second-nearest-neighbor interactions are included.

In Fig.~\ref{fig11}(a) we show the $l$-dependence of the scattered x-ray
intensity for close-packed structures with second neighbor interaction,
following the calculations by Hendricks and
Teller.~\cite{Hendricks42a,Kakinoki54a} Stacking faults can be introduced
by tuning the probability for second neighbor layers to be alike or
unlike. For details, see section~6 of Ref.~\onlinecite{Hendricks42a}. The
curves in Fig.~\ref{fig11}(a) were calculated for different degrees of
stacking disorder $z$, ranging from a predominant 2-layer period for $z\ll
1$ (central peaks at integer $l$) to a predominant 3-layer period for
$z\gg 1$ (split peaks at $l \pm 2/3$ with $l$ even). The dashed line (red)
line is for complete disorder ($z=1$), which is characterized by a central
peak for odd $l$ and no obvious remains of the peaks at even
$l$.~\cite{evenpeaks} In Fig.~\ref{fig11}(b) we show that, as the volume
fraction with a 3-layer period decreases, the splitting in $l$ decreases
and eventually disappears close to the point where complete disorder is
reached.

\begin{figure}[t]
\center{\includegraphics[width=0.85\columnwidth,angle=0,clip]{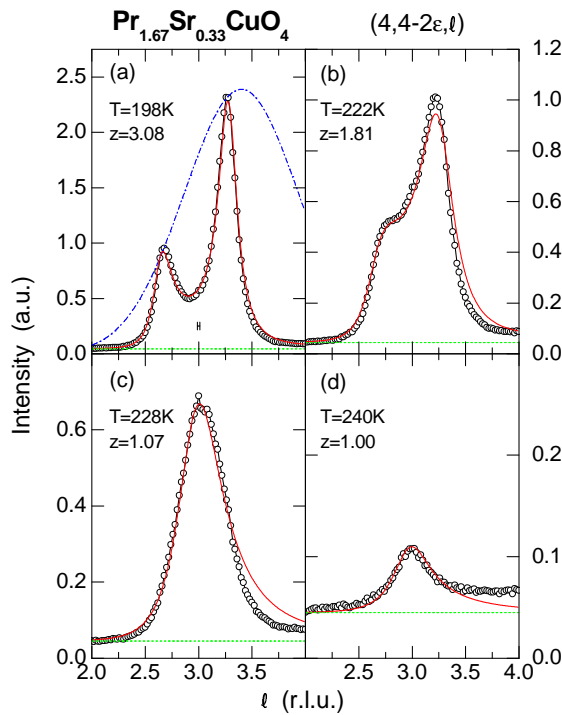}}
\caption[]{(Color online) $l$-scans through $(4,4-2\epsilon,3)$ charge
peak of Ar-annealed \psno\ plus fits (red solid line) as described in the
text. The dash-dot line (blue) represents the structure factor, the dotted
line (green) a $T$-independent constant background. Instrumental
resolution full width is indicated by H-sign in (a).} \label{fig12}
\end{figure}

The experimental results for Ar-annealed \psno , in Fig.~\ref{fig9},
clearly show that for $p=1/3$ a ground state with a 3-layer stacking
period is favored over a 2-layer period. First, this indicates that
stripes prefer to be all of the same type, i.e., either bond- or
site-centered, but not mixed as in Fig.~\ref{fig10}(c). Second, it shows
that the Coulomb repulsion between stripes in second neighbor planes is
significant. The fact that in Fig.~\ref{fig9}(a) at 10~K the splitting in
$l$ is somewhat smaller than $2/3$, indicates a certain degree of disorder
[cf. Fig.~\ref{fig11}(b)]. Furthermore, it is obvious that the peak
intensity at approximately $(4,4-2\epsilon,2+2/3)$ is significantly lower
than at $(4,4-2\epsilon,4-2/3)$. We assume that this difference is largely
due to the structure factor, since the Bragg intensity at (442) is about a
factor forty smaller than at (444).~\cite{ABCimbalance} From the width of
the larger of the two peaks we have estimated a correlation length of
$\xi_c=17.5$~\AA , which corresponds to approximately three $\rm NiO_2$
layers.~\cite{correlationlength}

In Fig.~\ref{fig12} we present fits to the data obtained by the matrix
method used by Hendricks and Teller,~\cite{Hendricks42a} multiplied by a
slowly varying structure factor $F^2$ for which we assume the
phenomenological, Gaussian $l$-dependence indicated in
Fig.~\ref{fig12}(a). The only variables are the stacking parameter $z$ and
the amplitude. A finite background estimated from $k$-scans, a small
offset in $l$ of -0.04 (explanation in Sec.~\ref{stripestacking}), and the
structure factor were kept constant for all temperatures. Obviously, the
model provides a good description of the data. Moreover, the parameter $z$
gives us access to the temperature dependence of the stacking disorder.
Note that fits with a different $F^2$ than the one shown in
Fig.~\ref{fig12}(a), e.g., with the maximum centered at $l=4$, had
virtually no effect on the $T$-dependence of $z$.

\subsection{The charge stripe melting process}
\begin{figure}[t]
\center{\includegraphics[width=0.75\columnwidth,angle=0,clip]{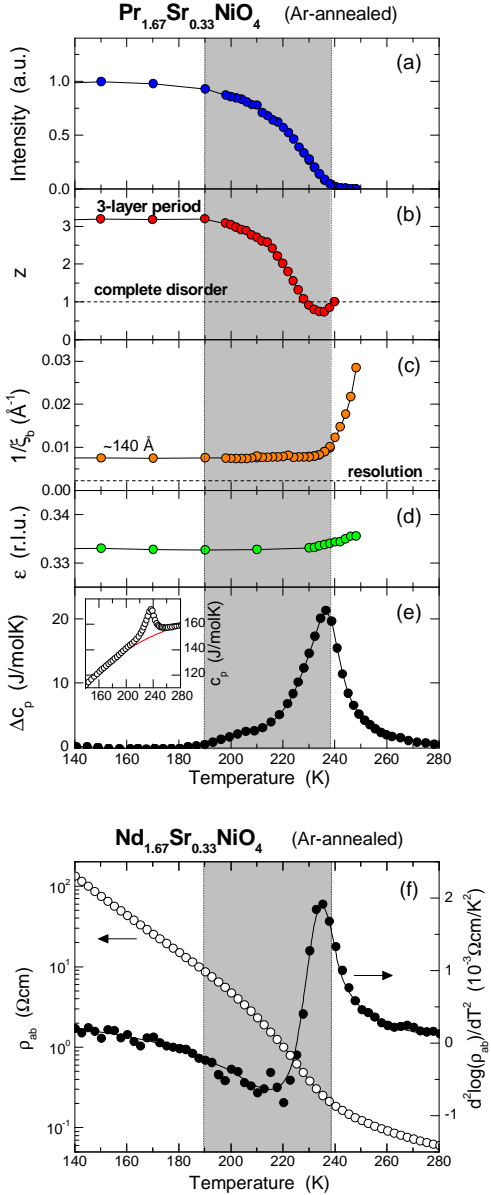}}
\caption[]{(Color online) Top: temperature dependence of various
properties of Ar-annealed \psno\ close to the charge stripe order
transition. (a) $l$-integrated intensity of ($4,4-2\epsilon ,3$) charge
peak, (b) parameter $z$ from fits of $l$-scans (see Fig.~\ref{fig12}), (c)
inverse in-plane correlation length from $k$-scans (d) incommensurability
$\epsilon$ (see Fig.~\ref{fig2}), (e) measured specific heat (inset) and
specific heat anomaly obtained by subtraction of 4th-order polynomial;
solid line (red) in inset. Bottom: $ab$-plane resistivity and its second
derivative of the Ar-annealed \nsno\ crystal.} \label{fig13}
\end{figure}

In Fig.~\ref{fig13} we compare the fitted $z(T)$ with various properties
of the Ar-annealed \psno\ and \nsno\ crystals, such as the $l$-integrated
charge-peak intensity, the inverse in-plane correlation length
$\xi_b^{-1}$ of the charge stripe order, the incommensurability
$\epsilon$, the specific heat $c_p$, and the in-plane resistivity
$\rho_{ab}$. A pronounced maximum is observed in $c_p$ at 238~K, which we
identify with $T_{\rm CO}$, the onset of static charge stripe order. Above
this temperature, charge peaks become very small, and $\xi_b^{-1}$ starts
to diverge, i.e., the in-plane stripe order melts.~\cite{Du00aN} In
contrast, the parameter $z$ already starts to decrease at $\sim$200~K and
reaches complete disorder ($z=1$) at about 228~K. From this we conclude
that the melting of the 3D charge stripe lattice sets in with a melting of
the interlayer correlations, leaving the 2D correlations intact, until
these eventually melt at $T_{\rm CO}$ as well.

In Fig.~\ref{fig13}(b) one can see that $z$ goes through a minimum and
then increases for $T\gtrsim 235$~K. This anew increase goes along with
the increase of $\xi_b^{-1}$ and reflects the drastic decrease of the
stripe correlations above \tco . Note that, in \lsno , a clear minimum in
$\xi_c^{-1}$ was observed at about the same temperature.~\cite{Du00aN} We
believe that there the increase of $\xi_c^{-1}$ at temperatures below the
minimum is a precursor of the formation of the 3-layer period stacking
order. According to Fig.~\ref{fig13}(b), $z<1$ at the temperature of the
minimum, which, if taken serious, implies a slight tendency towards a
2-layer stacking period. It is certainly possible that, once the stacking
order starts to melt, stripes can arrange more freely, and may assume a
body-centered stacking order, for which the ordering wave vector ${\bf
g}_{\rm CO}$ is the same as in the case of disorder [cf.
Figs.~\ref{fig8}(b) and \ref{fig10}(c)]. However, further work is
necessary to verify if our analysis, which is based on the model by
Hendricks and Teller, correctly describes even such small details.

The spin stripe ordering temperature $T_{\rm SO}$ of our crystals has not
yet been determined. While in \lsno\ both \tco\ and \tso\ can be
identified by weak anomalies in the static magnetic \sus\ of the $\rm
NiO_2$ planes, this is impossible in the case of our crystals, because of
the large paramagnetic contribution of the $\rm Pr^{3+}$ and $\rm Nd^{3+}$
ions.~\cite{Klingeler05aN} However, results from neutron diffraction in
Ref.~\onlinecite{Kajimoto03aN} show that, in \nsno , spin stripes order at
essentially the same temperature as in \lsno , with $T_{\rm SO}\simeq
190$~K. This means that the spin stripe order disappears just about the
temperature where the $c$-axis stacking correlations of the charge stripes
start to melt. In this context it is worth noting that the low temperature
side of the anomaly in $c_p$ has a shoulder which extends down to $T_{\rm
SO}$. Measurements of $c_p$ in Ref.~\onlinecite{Klingeler05aN} for a
\lsno\ single crystal show this even more clearly. We believe that the
shoulder indicates that the melting of the stripe stacking order is
associated with a significant entropy change.~\cite{entropy}

Fingerprints of the melting of the interlayer stripe order well below
$T_{\rm CO}$ are also observed in the electronic transport. In \nsno , in
Fig.~\ref{fig1}(f), the onset of the charge stripe order at $T_{\rm CO}$
is announced by a kink in log($\rho_{ab}$), as is also evident from a peak
in $d^2{\rm log}(\rho_{ab})/dT^2$. However, the entire transition in
log($\rho_{ab}$) is not sharp, but extends down to about 200~K, similar to
the behavior of $z$, $\Delta c_p$, and the charge-peak intensity.
One might think of this as a manifestation of the intimate connection
between charge and spin stripe order, since it implies that the charge
order has to be fairly progressed, before spin stripe order can occur.

Another interesting feature of the anomaly in $c_p$ is its extension
towards high temperatures. In this region charge peaks become extremely
weak and broad. It is reasonable to consider that both effects are related
to slowly fluctuating charge stripe correlations.~\cite{Du00aN} Intensity
from these correlations is picked-up due to the poor energy resolution of
the x-ray diffraction experiment. An inelastic neutron scattering study of
the incommensurable magnetic fluctuations in \lstno\ gives indirect
evidence for a liquid phase of charge stripes above $T_{\rm
CO}$.~\cite{Lee02aN}

The present observations are consistent with a stripe smectic or nematic
phase at $T>T_{\rm CO}$.~\cite{Kivelson98} (Note that this assignment is
different from that originally made by Lee and Cheong.~\cite{Lee97aN})


\subsection{Comparison with TEM data}
Our results concerning the stacking of stripes are in qualitative
agreement with a recent transmission-electron-microscopy (TEM) study on
tetragonal \lstno\ with x=0.275.~\cite{Li03aN} In this study, it was shown
that, in domains of mesoscopic dimension, charge stripes are indeed one
dimensional. However, the Sr content of $x=0.275$ gave rise to a mixing of
features expected either for $x=0.25$ or $x=0.33$, as well as features
unique to this intermediate doping. In particular, the obtained average
in-plane stripe distance ($1.75\times b$) results in a mixture of site-
and bond-centered stripes within the same $\rm NiO_2$ plane. As a
consequence, the stacking of stripes along $c$ is strongly disordered,
with both simple body-centered arrangements as well as staggered shifts by
$0.5 \times b$ prevalent.~\cite{Li03aN}

Here, we have shown that the orthorhombic symmetry of \psno\ and \nsno\
can be used as a tool to obtain large single crystals with unidirectional
stripe order, suited to the study of macroscopic properties. For $x=0.33$
and $\delta = 0$, stripes in a single $\rm NiO_2$ plane are likely to be
either all bond-centered or all site-centered, since the stripe distance
of $1.5\times b$ is commensurate with the lattice. In the case of long
range in-plane order, the Coulomb interaction between stripes in first and
second neighbor planes becomes crucial and stabilizes a stacking such as
depicted in Fig.~\ref{fig10}(a). Thus, for the first time, a stripe order
with a 3-layer period has been observed in the x-ray diffraction pattern
of a macroscopic crystal.

It is worthwhile mentioning that there are several studies of \lsno\ with
$\epsilon$ matching the critical value of 1/3 quite
well.~\cite{Lee97aN,Kajimoto01aN,Du00aN,Ishizaka04aN} However, a stripe
stacking order of the kind identified here in \psno\ and \nsno\ has not
been reported. These observations suggest that the stabilization of the
3-layer stacking period may benefit from the fact that, in orthorhombic
crystals, the in-plane stripe order is unidirectional.

\subsection{Comparison with \lscob }
Besides the obvious differences, the electronic phase diagrams of
nickelates and cuprates show some interesting similarities. In the present
context it is particularly remarkable that, in \lsco , as long as $x$ is
below the MI transition, a static diagonal spin stripe order
forms.~\cite{Wakimoto99a,Wakimoto00a,Fujita02a} It remains an open
experimental question, whether corresponding samples also exhibit diagonal
charge stripes. In this respect, it is an interesting finding that, in
nickelates with the LTO structure, $charge$ stripes prefer to be parallel
to the octahedral tilt axis, similar to the $spin$ stripes in the
cuprates. These findings thus provide motivation for further experimental
searches for charge stripes in the insulating cuprates.

There are certainly alternative models for the lightly-doped cuprates that
do not involve stripes ~\cite{Berciu04a,Sushkov04a,
Luescher05a,Lindgard05a,Gooding97}; however, there is also additional
evidence for the cuprates suggesting a common coupling of charge and spin
stripes to an orthorhombic lattice distortion. In \lcod\ and \lstco , both
systems with parallel stripes and LTO structure, it was observed that the
direction of the spin stripes is not perfectly parallel to the Cu-O bonds,
but slightly inclined towards the diagonally running octahedral tilt axis
($\parallel a$), by an angle too large to be explained with the
orthorhombic lattice distortion.~\cite{Lee99,Kimura00a} Unfortunately, in
these compounds, no charge peaks were detected. However, in
Ref.~\onlinecite{Kimura04a} and Ref.~\onlinecite{Fujita02b} it was shown
that in orthorhombic \lbsco\ with $x=0.075$, both spin stripes and charge
stripes show the same inclination towards the octahedral tilt axis. This
strongly suggests that, in both classes of materials, nickelates and
cuprates, the response of charge stripes to orthorhombic lattice
distortions is similar.

\section{Conclusion}
We have presented a detailed x-ray diffraction study which sheds light on
the coupling between the charge stripes and structural distortions in
\psno\ and \nsno\ single crystals. In contrast to the sister compound
\lsno , both crystals undergo a HTT$\rightarrow$LTO transition well above
room temperature, so that stripe order at \tco\ takes place in an
anisotropic environment. We find that the orthorhombic strain causes the
stripes to align parallel to the short $a$-axis, which is also the
direction of the $\rm NiO_6$ octahedral tilt axis. In addition to these
in-plane correlations, we have observed correlations between the planes,
which are consistent with a stacking period of three $\rm NiO_2$ layers.
This stacking order is extremely vulnerable to interstitial oxygen
impurities and deviations of the total hole concentration from $p=1/3$; as
a result, it was observed in Ar-annealed samples, only. Further, we find
that the melting of the static charge stripe order is a two-step process,
which starts at 200~K with the melting of the interlayer correlations, and
is completed at \tco\ with the melting of the in-plane correlations.
Implications for the stripe order in insulating \lsco\ have been
discussed. The observation of unidirectional stripes in \psno\ and \nsno\
opens up the unique possibility to characterize their
anisotropic properties.\\

The work at Brookhaven was supported by the Office of Science, U.S.
Department of Energy under Contract No.\ DE-AC02-98CH10886.


\begin{thebibliography}{10}
\bibitem{NSNOpeaks}
These reflections were measured after the \nsno\ crystal was exposed to
air for
  about one year. Therefore, the crystal contains some excess oxygen, resulting
  in a larger and more inhomogeneous hole concentration than immediately after
  the Ar-annealing. This is apparent from the larger value of $\epsilon$, and
  an approximately twice as large peak width [cf. Figs.~\ref{fig6}(d) and
  \ref{fig7}(b)].
\bibitem{LTLOpeak}
Note that we have integrated the intensity of the (330) peaks in both
domains, i.e., $A$ and $B$.
\bibitem{newXTAL}
We mention that a newly grown \nsno\ crystal shows a sharp LTO/LTLO
transition at $\sim$100~K and no intensity above that temperature. Results
will be published elsewhere.
\bibitem{evenpeaks}
Note that the Hendricks-Teller model allows us to simulate the mixing of
  3-layer and 2-layer period stacking segments at any ratio. In the case of a
  2-layer stacking period like the one depicted in Fig.~\ref{fig10}(b), peaks
  for $l$ even are allowed. However, it is very unlikely that in our samples
  the mixing of the 3-layer ground state with this 2-layer stacking type goes
  beyond the point of complete disorder. On the other hand, a ground state with
  a 2-layer period of the kind depicted in Fig.~\ref{fig10}(c) is possible. For
  this body centered stacking type, charge peaks for $l$ even should be absent.
\bibitem{BCBC}
Equivalent with this are the sequences ...BCBC... and ...CACA... .
\bibitem{ABCimbalance}
In Ref.~\onlinecite{Hendricks42a} it is mentioned that an imbalance in the
  population of domains with stacking sequence ...ABCABC... and ...CBACBA...
  can cause a similar effect. However, this effect should vary between
  different crystals. In contrast, in about half a dozen samples with a
  pronounced 3-layer stacking period, we always have observed a similar
  intensity ratio.
\bibitem{correlationlength}
The correlation lengths were calculated from the peak width using
$\xi_b=({\rm
  HWHM} \times b^*)^{-1}$ and $\xi_c=({\rm HWHM} \times c^*)^{-1}$ where HWHM
  is the half width at half maximum in reciprocal lattice units of
  $b^*=1.162~{\rm \AA} ^{-1}$ and $c^*=0.504~{\rm \AA} ^{-1}$, respectively.
\bibitem{entropy}
As was discussed in Ref.~\onlinecite{Klingeler05aN}, it is likely that
both spin as well as charge degrees of freedom contribute to the entropy.
\end{thebibliography}

\end{document}